# Spontaneous off-stoichiometry as the knob to control dielectric properties of gapped metals

Muhammad Rizwan Khan[1], Harshan Reddy Gopidi[1], Hamid Reza Darabian[1], Dorota A. Pawlak[1], and Oleksandr I. Malyi[1,#]

[1]Centre of Excellence ENSEMBLE3 Sp. z o. o., Wolczynska Str. 133, 01-919, Warsaw, Poland

[#]**Email:** oleksandrmalyi@gmail.com (O.I.M)

**Abstract:**
Using the first-principles calculations and $La_3Te_4$ as an example of an n-type gapped metal, we demonstrate that gapped metals can develop spontaneous defect formation resulting in off-stoichiometric compounds. Importantly, these compounds have different free carrier concentrations and can be realized by optimizing synthesis conditions. The ability to manipulate the free carrier concentration allows to tailor intraband and interband transitions, thus controlling the optoelectronic properties of materials in general. Specifically, by realizing different off-stochiometric $La_{3-x}Te_4$ compounds, it is possible to reach specific crossings of the real part of the dielectric function with the zero line, reduce plasma frequency contribution to absorption spectra, or, more generally, induce metal-to-insulator transition. This is particularly important in the context of optoelectronic, plasmonic, and epsilon-near-zero materials, as it enables materials design with a target functionality. While this work is limited to the specific gapped metal, we demonstrate that the fundamental physics is transferable to other gapped metals and can be generally used to design a wide class of new optoelectronic/plasmonic materials.

***Dielectric function as the response of a material to the electromagnetic field:*** Traditional solid-state physics books[1, 2] teach us that light-matter interaction is defined by the electronic and ionic response to the electromagnetic field and is characterized by a complex frequency ($\omega$)-dependent dielectric function, $\varepsilon(\omega)=\varepsilon_1(\omega)+i\varepsilon_2(\omega)$. The real part of the dielectric function ($\varepsilon_1(\omega)$) describes the ability of a material to polarize in response to an electric field, while the imaginary part of the dielectric function ($\varepsilon_2(\omega)$) is related to the absorption coefficient, measuring how much energy is absorbed per unit distance traveled through a given material. These quantities are related to each other via the Kramers-Kronig transformation[3, 4] and can be used to calculate other dielectric properties, such as the refractive index and extinction coefficient, as well as classifying different optoelectronic/plasmonic materials. For instance, solar cell absorber materials (e.g., halide perovskites) have a high imaginary part of the dielectric function to absorb a sufficient amount of light. In contrast, plasmonic materials (e.g., gold and silver[5]) have a low real part of the dielectric function in the visible and near-infrared region. Understanding the dielectric function of materials is also vital for photonics and metamaterials, where the manipulation of light-matter interactions is used to achieve desired outcomes, e.g., high-efficiency energy conversion or enhanced sensing capabilities.

***Critical needs for the knob to control dielectric properties of the materials:*** The properties of any given material are defined by atomic identities, composition, and structure (ACS)[6]. This implies that the design of materials for a target functionality requires identifying specific ACS accommodating specific material properties, and unless such materials are identified, the optimal device performance cannot be reached. For instance, for solar cells, a difference in band gap energy of 0.1 eV can be essential to change the theoretical Shockley–Queisser limit efficiency[7] by a few percent. Such robust functionality dependence on ACS makes identifying a knob to tune given properties (e.g., dielectric properties or band gap) the crucial task of material design. Consider a wide class of plasmonic materials where the dielectric properties are closely related to the screened plasma frequency, defined as the frequency at which $\varepsilon_1(\omega)$ is equal to zero[8]. By adjusting the knob, the screened plasma frequency can be tuned to optimize the material performance for a particular application, e.g., by shifting the wavelength range where the material behaves as a plasmonic material. For example, in modern semiconductor technologies, successful tailoring of the dielectric properties can lead to the development of more efficient transparent conductors[9]. In optical sensor technology, governing the dielectric properties of plasmonic materials allows to engineer more sensitive and selective sensors by enabling a stronger material response to given chemicals or biological molecules[10]. The critical need for such a knob becomes even stronger with the discovery of Epsilon-Near-Zero (ENZ) materials[11], offering the capability of ENZ frequency tuning under which the real part of permittivity approaches near zero value.

***Gapped metals as examples of optoelectronic and plasmonic materials:*** Due to high electron density and a large surface plasmon resonance, metals with a continuous electronic density of states are considered to be classical plasmonic materials[5, 12]. Properties of such compounds are typically difficult to control and manipulate without the use of external factors (e.g., doping, pressure). However, with the development of electronic structure theory, it became clear that a continuous electronic structure does not always characterize metallic properties. Indeed, there is a range of so-called n-type gapped metals (p-type gapped metals also exist but are rarer[13, 14]) that, in undoped form, have a finite energy gap between their principal band edges with Fermi level in the principal conduction band (Fig. 1a). The electronic structure of such pristine gapped metals resembles that of degenerate semiconductors, received by heavy doping of wide band gap compounds (e.g., Al-doped ZnO[15, 16]) with the difference that gapped metals are received without any intentional doping and have carrier concentration significantly higher than degenerate semiconductors. The noticeable examples of such gapped metals include: $SrVO_3$ and $CaVO_3$ - "correlated" metals being potential transparent conductors[17, 18], $Sr_{1-x}NbO_3$ - color metallic photocatalyst[19], $Ca_6Al_7O_{14}$ - metallic solid-state electride (material where electrons are present as anions in the crystal structure, rather than being associated with particular atoms)[20], $La_3Te_4$ - an example of intrinsic thermoelectrics[21], and $SrVO_3$, $CaNbO_3$, $SrNbO_3$, and $BaNbO_3$ - promising plasmonic materials [22, 23].

***Spontaneous off-stoichiometry in gapped metals - $La_3Te_4$ as n-type gapped metal that can be synthesized in different stoichiometries:*** What makes the case of gapped metal special is that these compounds can exhibit the spontaneous formation of a point defect inducing noticeable change material stoichiometry. Specifically, the formation energy of a given defect becomes negative for a given chemical potential (synthesis conditions, e.g., $\mu = \mu° + RT \ln(P/P°)$, where $\mu°$ is the standard chemical potential, R is the universal gas constant, T is the temperature, and P and P° are the pressure of the gas and the reference pressure, respectively). It is important to distinguish this phenomenon from that of regular solid compounds, where the formation of point defects is typically limited until high temperatures and is driven by the minimization of Gibbs's free energy via increasing entropy at a given temperature. In contrast, the formation of an acceptor defect in n-type gapped metal (Fig. 1a) can develop off-stoichiometry due to the decay of conduction electrons to the acceptor level (known as Fermi level instability)[9, 14, 24]. Such behavior can lead to the formation of a range of non-stoichiometric compounds due to the intrinsic non-thermal tendency of the system, even at low temperatures. For instance, previous experimental works[25, 26] and our recent first-principles calculations[24] showed that $La_3Te_4$ ($I\bar{4}3d$, No. 220) is the example of n-type gapped metal (Fig. 1a,b) that can develop such non-stoichiometry as a result of the formation of acceptor La vacancies. Moreover, upon reaching x= 1/3 in $La_{3-x}Te_4$, all free carriers are removed from the principal conduction band (i.e., the system becomes an insulator), limiting the spontaneous defect formation – the formation energy of La vacancy in $La_{2.67}Te_4$ is positive. The findings are consistent with experimental crystallographic data[27] and can be explained by removing 3e from the principal conduction band with the formation of each La vacancy. We emphasize that the origin of such spontaneous defect formation is Fermi level instability of gapped metal compound and is not driven by temperature effect or stabilization of metastable phases. When all electrons from the conduction band are removed, the system cannot further benefit from electron-hole recombination. Hence, an increase in off-stoichiometry is not thermodynamically favorable. To demonstrate this, we build an energy convex hull[28] using calculated formation heats of various La-Te phases. By definition, compounds that do not tend to decompose spontaneously to competing phases have 0 meV/atom energy above hull ($E_{hull}$). In contrast, compounds above convex hull have positive $E_{hull}$ corresponding to exothermic energy of decomposition with respect to competing phases (we note that realization of compound <100-150 meV/atom is often still possible).[29, 30] All these compounds have different free carrier concentrations but the same parental structure (i.e., one can be considered as $La_3Te_4$ ($I\bar{4}3d$) structure decorated by La vacancies), which is in line with available experimental data[25-27, 31]. Hence, tunning potential synthesis conditions can be used to stabilize target composition with desired electronic properties and free carrier concentration. These results agree with experimentally observed results showing how controllable resistivity and even metal-to-insulator transition can be reached via optimized synthesis conditions.[26, 27]

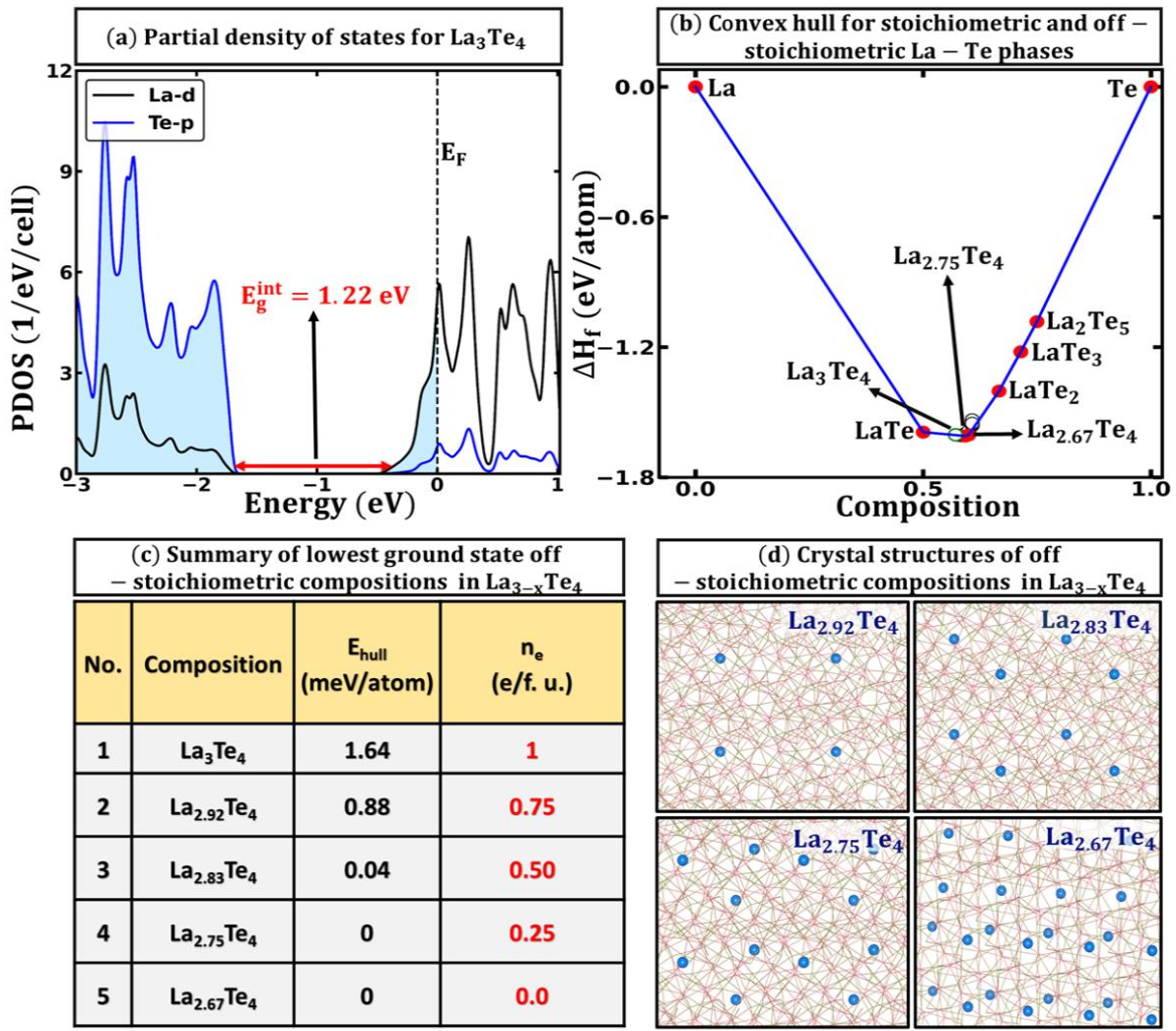

**Figure 1:** Spontaneous off-stoichiometry in La$_3$Te$_4$ n-type gapped metal. (a) Calculated projected density of states for La$_3$Te$_4$. (b) Calculated energy convex hull for the La-Te systems, including all known experimentally reported stoichiometric compositions and generated off-stoichiometric structures (only the lowest energy configurations for each off-stoichiometric composition shown are visualized). (c) List of generated off-stoichiometric composition and the corresponding energy above convex hull (E$_{hull}$) and free carrier concentration (n$_e$). (d) Crystal structures of off-stoichiometric compositions La$_{2.92}$Te$_4$, La$_{2.83}$Te$_4$, La$_{2.75}$Te$_4$, and La$_{2.67}$Te$_4$, with blue circles depicting the vacancies.

***Dielectric properties of La$_3$Te$_4$ defined by superposition of band-to-band transitions and Drude contribution:*** Similar to regular metals, dielectric properties of gapped metals are defined by superposition of band-to-band transition and free carrier absorption (Drude term) as schematically shown in Fig. 2a. One should note, however, that in the case of gapped La$_3$Te$_4$ metal, there are two band-to-band transitions in the system: (i) transition from occupied states in the principal conduction band to above unoccupied states and (ii) transition from an occupied state of principal valance band to unoccupied states in the conduction band. We note that these transitions are activated at different energies, which thus makes gapped metals substantially different from traditional metals having continuous electronic structures. At frequencies comparable to plasma frequencies $\omega_p$, dielectric properties of metals have a noticeable free carrier contribution as $\varepsilon_2(\omega) = \gamma \omega_p^2 / \omega(\omega^2 + \gamma^2)$ and

$\varepsilon_1(\omega) = 1 - \omega_p^2/(\omega^2 + \gamma^2)$, where $\gamma$ is a damping coefficient characterizing the average time interval between successive collisions experienced by a charged particle moving through a material[32] and $\omega_p$ is unscreened plasma frequency (see methods). The results shown in Fig. 2c,d indicate that the Drude contribution to the imaginary part of the dielectric function is positive and tends to decrease as the frequency increases, eventually approaching a zero limit in the infinite frequency range. In contrast, the Drude contribution to the real part of dielectric function is negative for $\omega_p \gg \omega$ and approaches 1 limit with increased frequency. The superposition of the Drude contribution and band-to-band transition results in a non-monotonic profile of real and imaginary parts of dielectric functions (Fig. 2c,d). For instance, a pristine $La_3Te_4$ compound has multiple crossings of the real part of the dielectric function with zero line, which potentially can make this material interesting for plasmonic and even ENZ materials (Fig. 2c). The imaginary part of the dielectric function and absorption spectra have noticeable Drude contribution at low frequencies, which become a non-dominant factor at high frequencies. Generally, the dielectric properties of $La_3Te_4$ are very similar to that in potential intrinsic transparent conductors[9], with only the difference that $La_3Te_4$ is not transparent - i.e., it has strong absorption in the visible light range.

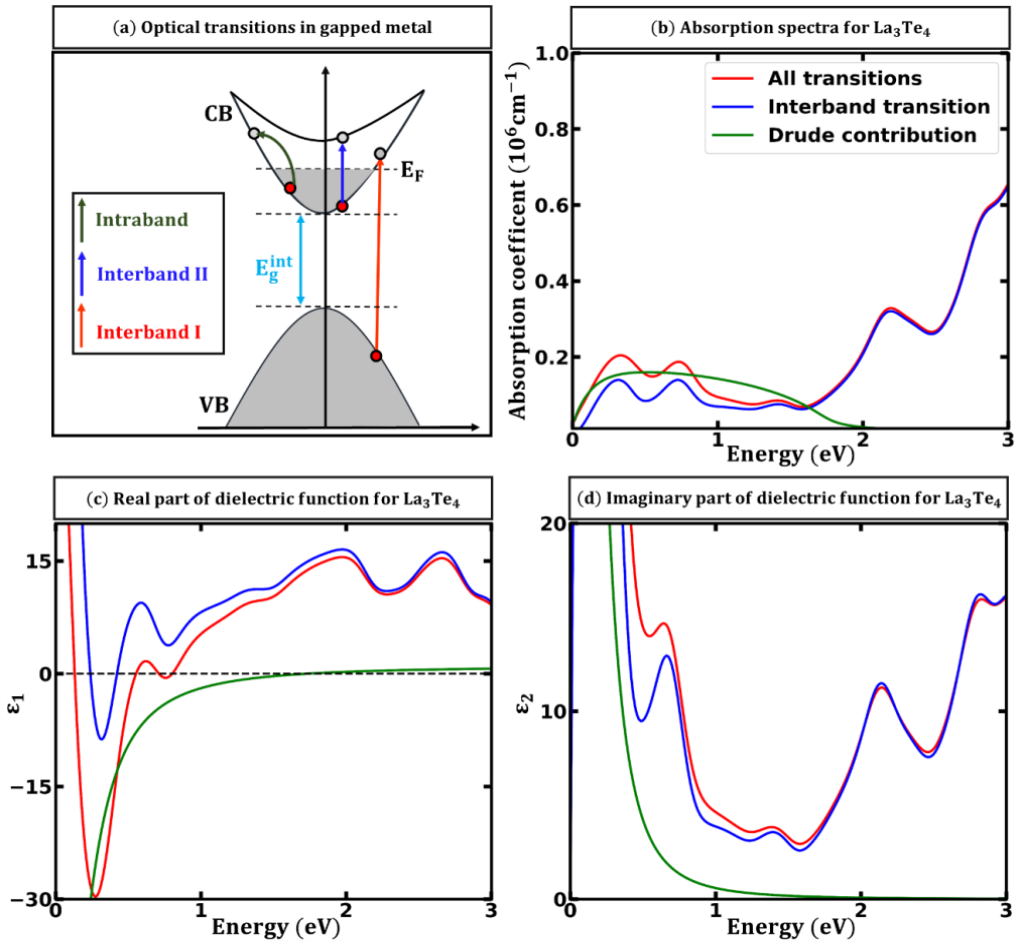

**Figure 2: Dielectric properties of $La_3Te_4$ as superposition of band-to-band transitions and Drude contribution.** (a) Schematic figure demonstrating different types of optical transition present in the gapped metal. (b) Absorption spectra for $La_3Te_4$ considering only interband transitions, Drude contribution, and superposition of interband and intraband transitions. (c) Real and (d) imaginary parts of dielectric function for $La_3Te_4$.

***Spontaneous off-stoichiometry as a knob to tune the optoelectronic properties of $La_{3-x}Te_4$ system:*** As noted in Fig. 1c, different off-stoichiometric compounds have different numbers of free carriers. Taking into account that unscreened plasma frequency ($\omega_p$) is directly defined by free carrier concentration

(in the simplified textbook picture, $\omega_p \sim \sqrt{\frac{n_e}{m}}$, where $n_e$ and $m$ are free carrier density and effective mass, respectively), the La deficiency directly reduces the plasma frequency via the reduction of carrier concentration as shown in Fig. 3a. Here, one should note that due to the distribution of vacancy sides, there is a small variation of plasma frequency along different directions (within a couple of percent scale). However, this anisotropy would most probably be overcome with an increased temperature-driven disorder (e.g., entropy). The illustration of this case is well seen within the absorption spectra, where the reduction of plasma frequency induces a reduction of free carrier absorption at low frequencies. Another important effect seen in Fig. 3b and Fib. 3d is the red shift of the absorption spectra with increased off-stoichiometry. These results can be understood in the language of the Burstein-Moss effect[33, 34]. Specifically, due to the gapped metal electronic structure, the internal band gap of $La_3Te_4$ (1.22 eV) is smaller than the first available transition from occupied states in the valence band to unoccupied states in the conduction band (i.e., 1.74 eV for $La_3Te_4$) as also can be understood from schematic Fig. 2a. However, as the degree of off-stoichiometry (x) increases, fewer and fewer states are occupied in the conduction band reducing thus the energy of the first available direct transition. For instance, this energy is 1.59, 1.53, and 1.44 eV for $La_{2.92}Te_4$, $La_{2.83}Te_4$, and $La_{2.75}Te_4$, respectively. Eventually, at x=1/3, the system becomes an insulator (all free carriers are removed from the conduction band) with the band gap energy of 1.13 eV - no Burstein-Moss effect is present and the first possible direct transition corresponds to the lowest direct band gap energy. The degree of off-stoichiometry also has a noticeable effect on the real part of the dielectric function, which mainly can be summarized as a gradual reduction of Drude contribution resulting in a gradual redshift of a low-frequency crossing point of $\varepsilon_1(\omega)$ with zero line from about 0.7 eV to its complete suppression when at x=1/3. Taking into account the similarity of electronic structures for different $La_{3-x}Te_4$ compounds, we also can understand the origin of specific $\varepsilon_1(\omega)=0$ points. Specifically, the disappearing of $\varepsilon_1(\omega)$ crossing at low frequency is clearly attributed to the removal of conductive electrons from the principal conduction band as their removal lead to the disappearance of the crossing point. In contrast, for all $La_{3-x}Te_4$ compounds, $\varepsilon_1(\omega)=0$ at ~4.3 eV is clearly not affected by the presence of free carriers, implying that it is the intrinsic tendency of the system originated from band-to-band transitions (only one taken into account at these frequencies for all compounds). We note as well that generally the absolute values of crossing frequency should be taken into account with caution as the calculation presented here are done within the DFT framework resulting in the underestimation of the internal gap between principal band edges[24] or even sensitivity to scattering rate in the Drude model. Despite this, the demonstrated knob reflects the universal approach that can be adapted for a wide class of gapped metals to tailor their properties.

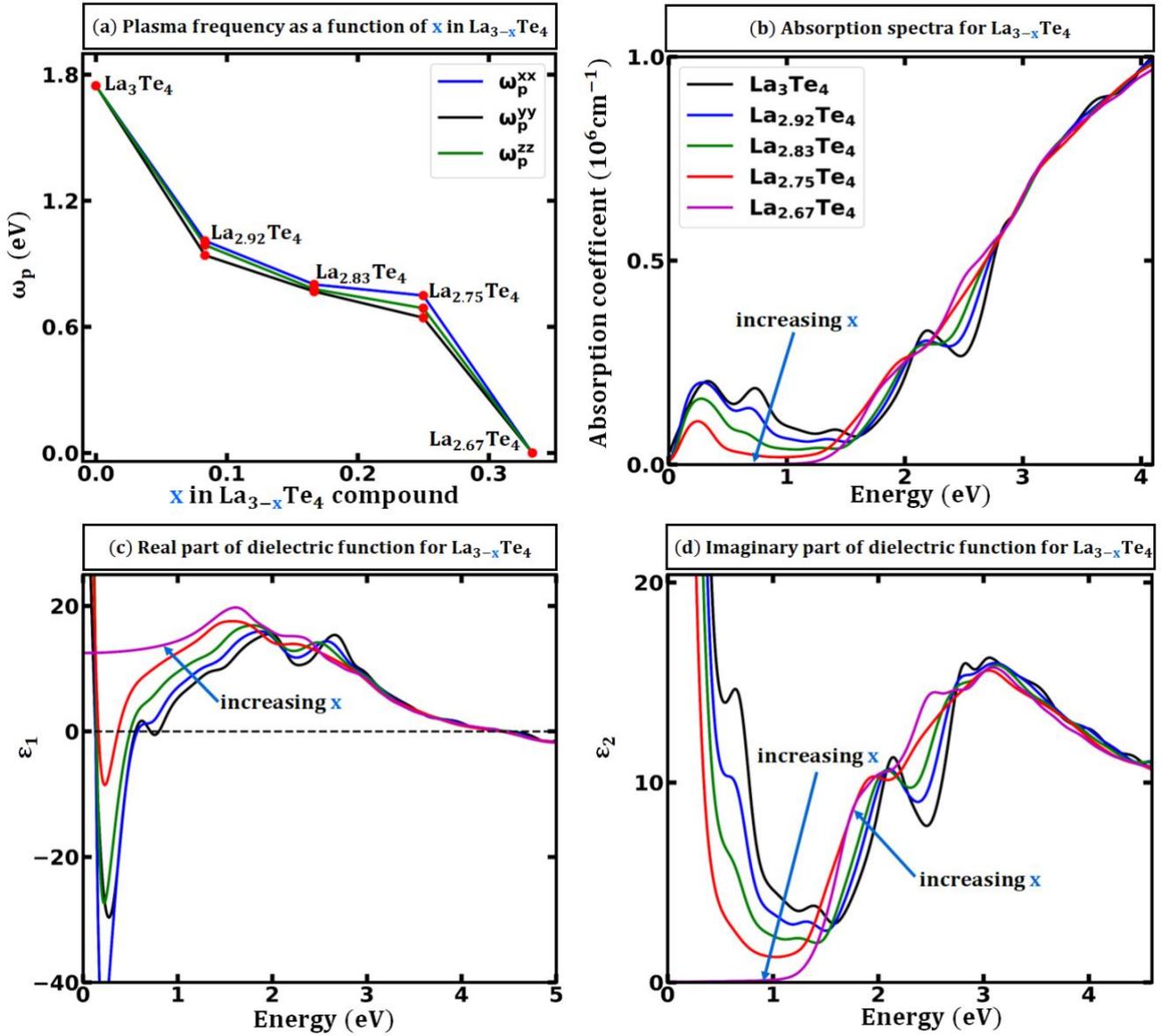

**Figure 3: Spontaneous off-stoichiometry as a knob to tune the optoelectronic properties of La$_{3-x}$Te$_4$ system.** (a) Plasma frequency as a function of different concentrations of La vacancies in La$_{3-x}$Te$_4$; (b) Absorption spectra for La$_{3-x}$Te$_4$ considering superposition of both interband and intraband transitions; (c) Real and (d) imaginary parts of dielectric function for different concentrations of La vacancies in La$_{3-x}$Te$_4$.

To summarize, using the example of La$_3$Te$_4$, we demonstrate that the representative example of n-type gapped metal can exhibit the spontaneous formation of a point defect (acceptor defect) due to the decay of conduction electrons to the acceptor level known as Fermi level instability even at low temperatures. This behavior is different from regular solid compounds, where the formation of point defects is typically limited until high temperatures and is driven by the minimization of Gibbs's free energy via increasing entropy. Importantly, spontaneous off-stoichiometry results in the formation of a set of non-stoichiometric compounds with the same parental crystal structure but different concentrations of free carriers. The formation of these compounds can be controlled through an optimized choice of chemical potentials (synthesis conditions), resulting in tunable optoelectronic properties. Specifically, owing to the gapped metallic electronic structure, controllable growth of off-stoichiometric La$_{3-x}$Te$_4$ compounds can be used to tailor the Drude contribution and optimize the band-to-band transition via controllable free carrier concentration. This is particularly important in the context of plasmonic and epsilon-near-zero materials, as it enables the development of new materials with a target functionality by using spontaneous off-stoichiometry as an external knob.

**Methods:**

**First-principle calculations:** We performed ab initio calculations using the Perdew-Becke-Ernzerhof (PBE) functional [35] as implemented in the Vienna Ab initio Simulation Package (VASP)[36-40]. The cutoff energy values for the plane wave basis were set to be 550 and 500 eV for static and final volume relaxation, respectively. The Γ-centered Monkhorst–Pack k-grids[41] scheme were used for Brillouin zone sampling with approximately 3000 and 10000 per reciprocal atom for volume relaxation and final static calculations, respectively. The results were analyzed using pymatgen[42] and Vesta[43].

**Generation of off-stoichiometric compounds and defect calculations:** To investigate the stability of off-stoichiometric $La_{3-x}Te_4$ compounds, we employed a supercell and generated five off-stoichiometric compositions by creating supercells containing 1, 2, 3, 4, and 5 La vacancies, respectively. We then conducted DFT calculations on each unique system resulting from at least 40 different randomly selected atomic configurations for each composition. To ensure comprehensive analysis, we also incorporated other theoretically known La-Te phases available in materials databases such Materials Project[44], Open Quantum Materials Database (OQMD)[45, 46], and Inorganic Crystal Structure Database (ICSD)[47]. The details on the defect calculations are given in Ref. [24].

**Calculations of the optical properties:** To compute the optical properties for gapped metals, first, we determine the frequency-dependent dielectric function ε(ω) based on density functional theory (DFT). This dielectric function consists of two terms, including direct interband transition matrix and term due to plasmonic effect based on Drude model[32]. The first term is calculated by considering only the optical interband transition while plasma frequency can be calculated from DFT electronic band structure, which is at q=0 plasmonic wave vector given as:

$$\omega_p^2 = ne^2/\pi^2 m^2 (\Sigma_a \int dk <\psi_{ak}|P^i|\psi_{ak}><\psi_{ak}|P^j|\psi_{ak}>\delta(E_{ak}-E_F)$$

Here, $\psi_{ak}$ and $E_{ak}$ are the wave function and energy of free electrons with charge(mass) e(m), respectively of band *a*, at momentum **k**. $E_F$ is correspond to the Fermi level and $P^i$ ($P^j$) are components of optical dipole transitions. Here, we mainly focus on the effects of direct band transition and intraband contributions in terms of plasma frequencies on optical properties by neglecting the small contributions of ions at low frequencies. To perform the calculations for direct band transitions and plasma frequencies, a Γ-centered Monkhorst–Pack k-grids[41] of 20000 per reciprocal atom were used along with a small complex shift (η) of 0.01 in the Kramers-Kronig transformation[48]. To include the Drude contribution term in the optical properties we utilized the kram code[49-51], with the plasma frequencies obtained from first principles calculation and the damping coefficient ($\gamma$) which is set to be 0.2 eV/ℏ, similar as used for another transparent conducting gapped metals and traditional doped transparent conductors[9, 52, 53].

**Acknowledgments:** The authors thank the "ENSEMBLE³ - Centre of Excellence for nanophotonics, advanced materials and novel crystal growth-based technologies" project (GA No. MAB/2020/14) carried out within the International Research Agendas programme of the Foundation for Polish Science co-financed by the European Union under the European Regional Development Fund and the European Union's Horizon 2020 research and innovation programme Teaming for Excellence (GA. No. 857543) for support of this work. We gratefully acknowledge Poland's high-performance computing infrastructure PLGrid (HPC Centers: ACK Cyfronet AGH) for providing computer facilities and support within computational grant no. PLG/2022/015458.